\documentclass[10pt]{article}
\usepackage{amssymb,amsmath,amscd,latexsym}
\begin{document}

\headsep -1.0cm
\title{On Adler's ``Conway--Kochen Twin Argument''}
\author{John Conway$^\star$ and Simon Kochen\thanks{jhorcon@yahoo.com; \
$^\star$kochen@math.princeton.edu}\\
Princeton University, Department of Mathematics\\ Princeton, NJ 08544-1000}

\maketitle

\newcommand{\ra}{\mathrm A}
\newcommand{\rb}{\mathrm B}
\newcommand{\iap}{\mathrm A'}
\newcommand{\ibp}{\mathrm B'}

\parskip .3cm

In arXiv: quant--ph/0604122v1 (17 April 2006), Stephen Adler published
a critique of our paper "The Free Will Theorem" (arXiv: quant--ph/0604079,
of which a slightly updated version [1] appears  in ``Foundations of
Physics'').  More recently we have discussed our argument with
Professor Adler, and are happy to say that he now agrees with us, after
certain clarifications, (see Adler [2]). 
Adler's main concern is that spin measurements
are not instantaneous, and also that the time they take varies with
direction. The information available to the particle will vary with this
time.  This is true, but causes no problem.  Particle a's  response
function $\theta_a$ may depend on all the  information that arrives up to
the time of its response.  We can in fact allow its domain to include
information up to a time T that maximizes this response time over the 40
triples used in our argument.  Of course $\theta_a$ for a particular triple
will not in fact vary with information that arrives  later than the
response time for that triple, but that does not affect its functional
nature (a function does not have to vary with all its arguments).  A
similar argument applies to $\theta_b$.

(In fact our argument took this into account, though only implicitly,
because the information $\alpha'$ of which $\theta_a$ is a function was
defined to be independent of $x,y,z$.   However, we are glad to provide
the extra clarification.)

In view of this, we do not actually need the ``sequence of
experiments'' that Adler goes on to discuss. Each of the two
experimenters of our argument makes just one experiment; for $\rb$ this is
to measure the squared spin in a direction $w$, while $\ra$ makes three such
measurements in orthogonal directions $x,y,z$.    The reason why no
sequence of such experiments is needed is that our ``functional
hypothesis`` threatens  to produce what we called a ``101--function'' of
direction.  Since no such function exists there must be a failure of one
of our axioms SPIN, TWIN or FIN. A failure of SPIN would be revealed if
A were to measure just the failing triple $x,y,z$, while a failure of TWIN
at $w$ would be revealed if $\rb$ were to measure $w$ and $\ra$ any 
triple including $w$. (As we said in the paper,  FIN is justified not 
by direct experiment but as a consequence of relativity.)

We are grateful to Adler for pointing us to the three papers [3],
[4], [5]. These also twin the K--S paradox to obtain ``no--go theorems''
that go beyond those in the book by Redhead referenced in our paper.
However, the Free Will Theorem is even stronger, in the ways mentioned
there.

We add some remarks on another topic. Several physicists have
proposed a type of model in which superluminal transmission is possible
in one sense (say, ``on the inner level''), but not in another 
(``on the outer level''). For instance, the hits of potential GRW 
explanations will be on the inner level, but experimenters cannot send messages
superluminally, since they can only access the outer level.  Indeed,
the Janus models of our paper are of this type, since Janus acts
superluminally, but they are not covariant, because the image of one
Janus model under a Lorentz transformation is another Janus model.  A
general strategy --- we  might call it the multi--Janus approach --- for
achieving covariance is to combine all these images in some way.  One
way of doing this was advocated in Fleming [6]; a similar one, based on
a paper of Aharanov and Albert [7],  was suggested for the GRW model by
Ghiradi [8].

These may indeed be covariant in a formal sense, but the Free Will
Theorem shows that any such strategy, if it works,  must have the
``dirty needles`` peculiarity described in the final version [1] of our paper,
which discusses theories that operate by injecting new information into
an otherwise deterministic universe. We show there that even if this new
information is intrinsically stochastic and/or non--locally correlated,
this will not help to account for our spin experiments, unless the
injection ``uses dirty needles''. By this we mean that either the
information injected near particle $b$ provides knowledge about
experimenter A's choice of $x,y,z$, or that injected near a involves
knowledge of $w$.

Each of these clearly contradicts the Free Will Assumption, since in a
suitable inertial frame  $b$'s response  happens before A chooses
$x,y,z$, while  in another frame $a$'s response happens before B chooses $w$.
Alternatively, we can view this as a violation of causality that is
gross in two ways --- because it concerns the location of spots on
macroscopic screens, and because the response of a particle can depend
on decisions that may only be made 5 minutes after that response.
However, these violations have a curious kind of deniability, in that
they are undetectable by observers who do not have access to all the
information involved.

To summarize; explanations of this two--level type face a dilemma ---
injections by ``clean needles'' cannot account for our spin experiments,
while the use of dirty needles is a gross violation of both causality
and the Free Will Assumption.


\begin{thebibliography}{9}

\bibitem[1]{1} J. H. Conway, S. Kochen,{\it Found. Phys.}
 (online July 11, 2006).

\bibitem[2]{2} S. Adler arXiv: quant--ph/0604122 (17 Oct. 2006).

\bibitem[3]{3} A. Stairs, {\it Phil. Sci.}
{\bf 50} (1983), 587.

\bibitem[4]{4} H.  R. Brown, G. Svetlichny, {\it Found. Phys.}
{\bf 20} (1990), 1379.

\bibitem[5]{5} A .Elby, {\it Found. Phys.}
{\bf 20} (1990), 1389.

\bibitem[6]{6} G. N. Fleming, {\it J. Math. Physics}, 
{\bf 7}, (1966), 1959.

\bibitem[7]{7} Y. Aharanov,  D. Z. Albert, {\it Phys. Rev.}
{\bf D 29} (1984), 228.

\bibitem[8]{8} G. Ghirardi, arXiv: quant--ph/0003149v1 (31 March 2000).
\end{thebibliography}
\end{document}